\documentclass[12pt]{article}

\begin{document}


\title{\large\bf Scalar-Tensor Gravity Theory For Dynamical Light
Velocity}

\author{ M.~A.~Clayton$^\dagger$
and J.~W.~Moffat${ }^*$\\ %
\\%
${^\dagger}$\textit{Department of Physics, Virginia Commonwealth
University,} \\ %
\textit{Richmond, Virginia 23284-2000,
 USA} \\%
\\%
${}^*$\textit{Department of Physics, University of Toronto,} \\ %
\textit{Toronto, Ontario M5S~1A7, Canada}}
\date{\today}

\maketitle

\begin{abstract}%
A gravity theory is developed with the metric ${\hat g}_{\mu\nu}=
{g}_{\mu\nu}+B\partial_\mu\phi\partial_\nu\phi$. In the present
universe the additional contribution from the scalar field in the
metric ${\hat g}_{\mu\nu}$ can generate an acceleration in the
expansion of the universe, without negative pressure and
with a zero cosmological constant. In this theory, gravitational waves will
propagate at a different speed from non-gravitational waves. It is
suggested that gravitational wave experiments could test this observational
signature.
\end{abstract}



\section{Introduction}

Recent observations of apparent luminosities of Type Ia supernovae (SNe-Ia)
at moderate redshift ($z\sim 0.6$) indicate that the universe is expanding
at an accelerated rate~\cite{Perlmutter}. If this observational
evidence is correct, then the implications for cosmology are
remarkable. Attempts to explain this phenomenon include
``quintesssence''~\cite{Steinhardt}, the cosmological
constant~\cite{Starobinsky}, a domain wall dominated
universe~\cite{Spergel} and non-perturbative vacuum contributions
to the effective action of a massive scalar field~\cite{Parker}.
In general it would appear that the cosmological fluid is
dominated in the present universe by an exotic energy density,
which has a negative pressure and which did not play an important
role in the early universe. The fine tuning problem related to
the cosmological constant is
well-known~\cite{Starobinsky,Weinberg} and amounts to a fine
tuning of order $10^{100}$ between the early universe
inflationary phase and the present universe. For the quintessence
model described by a slowly rolling scalar field, the potential
has to be extremely flat, so that it cannot roll to its true
minimum in the present universe. Characterizing the flatness by a
mass $m$ requires that the mass be extraordinarily ultra-light
$m\sim 10^{-33}$ eV. A mass of the same size is implied by the
non-perturbative vacuum driven mechanism.

All standard gravity theories, including Einstein's general relativity (GR),
are normally required to satisfy the positive energy conditions for
matter in the present universe, since over extended times this is perceived
as a reasonable physical requirement of a gravity theory. For the vacuum
energy, $p=-\rho$, which leads to a cosmological constant $\Lambda$. It is
believed that symmetry principles exist in particle physics which would
force $\Lambda$ to be zero, but there have been no convincing arguments
found to support this. How can we then describe the apparent speeding up of
the expansion of the universe in a gravity theory without violating the
positivity conditions on the density and pressure? Can we construct a
self-consistent gravity theory which can explain the data without violating
the positivity conditions and with a zero cosmological constant? In the
following, we shall develop such a gravity theory based on a
metric given by
\begin{equation}
\label{bimetric}
{\hat g}_{\mu\nu}=A[\phi]g_{\mu\nu}+B[\phi]\partial_\mu\phi\partial_\nu\phi,
\end{equation}
where $\phi$ is a scalar field and $\partial_\mu\phi=\partial\phi/\partial
x^{\mu}$. The inverse metrics ${\hat g}^{\mu\nu}$ and $g^{\mu\nu}$ satisfy
\begin{equation}
{\hat g}^{\mu\alpha}{\hat g}_{\nu\alpha}={\delta^\mu}_\nu,\quad g^{\mu\alpha}
g_{\nu\alpha}={\delta^\mu}_\nu.
\end{equation}
When $A[\phi]=1$ and $B[\phi]=0$, we retrieve conventional GR. As an
immediate simplification, we assume that $A[\phi]=1$ and $B[\phi]=B={\rm
constant}$, which serves to eliminate contributions to the field equations
for $\phi$ with more complicated dependence on derivatives of the source
tensor. The assumption that $A[\phi]=1$ is motivated by the fact that the,
essentially, conformal factor has been exhaustively studied in the
Brans-Dicke scenario~\cite{Dicke,Bertolami} and we are interested
in novel effects.

In a previous work~\cite{Vector}, a bimetric gravity theory was
constructed based on a metric similar to~(\ref{bimetric}) but in
which the second term was described by a vector field
$\psi_{\mu}$. This model provided a dynamical mechanism for a
superluminary theory in which light travels faster in the early
universe, thereby resolving problems in
cosmology~\cite{Vector,Superlum1,Albrecht,Superlum2}.

In the gravity theory presented in the following, we shall find that there
is an extra contribution to the gravity component in the equations of
motion for the expansion factor in cosmology. This contribution can lead to
an acceleration of the expansion of the universe.
The equation of state satisfies the standard positivity conditions for the
density and pressure. A fit to the type Ia supernovae (SNe-Ia) data is
obtained.

\section{The Action and Field Equations}

The model that we consider here is identical in spirit to that
which appeared in an earlier publication~\cite{Vector}, with the
important difference that the coupling is through a scalar field
which, given the predominance of scalar fields in cosmological
models~\cite{KT} and as effective models of more fundamental
theories such as string theory, is more in line with current
research in the field.  Here we will constrain ourselves to the
simplest of the class of such models so that we focus on novel
effects and avoid confusing the issue with Brans-Dicke-like and
dilaton couplings. These are easily included.

The model consists of three parts, represented by the three
separate contributions to the action:
\begin{equation}
S=S_{\rm grav}+S_{\phi}+S_{\rm M}.
\end{equation}
The standard general relativity contribution is:
\begin{equation}
S_{\rm grav}=-\frac{1}{\kappa}\int d^4x\sqrt{-g}(R[g]-2\Lambda),
\end{equation}
where $\kappa=16\pi G/c^4$, $\Lambda$ is the cosmological
constant, and we employ a metric with a ($+$$-$$-$$-$)
signature.  We also have a contribution from a minimally-coupled
scalar field:
\begin{equation}
S_{\rm \phi}=-\frac{1}{\kappa}\int d^4x\sqrt{-g}
\Bigl[\frac{1}{2}g^{\mu\nu}\partial_\mu\phi\partial_\nu\phi-V(\phi)\Bigr],
\end{equation}
where the scalar field $\phi$ has been chosen to be dimensionless.
The stress-energy tensor of the scalar field is of the standard
form:
\begin{equation}
\label{phiT}
T_\phi^{\mu\nu}=\frac{1}{\kappa}\Bigl[g^{\mu\alpha}g^{\nu\beta}\partial_\alpha\phi
\partial_\beta\phi-g^{\mu\nu}\Bigl(\frac{1}{2}g^{\alpha\beta}
\partial_\alpha\phi\partial_\beta\phi-V(\phi)\Bigr)\Bigr],
\end{equation}
and results from the variation of the scalar field action with
respect to $g_{\mu\nu}$.

Where our model departs from conventional model-building wisdom is
in the coupling of the gravitational field to matter.  Instead of
constructing the matter action using the metric $g_{\mu\nu}$, we
use the combination~(\ref{bimetric}), resulting in the
identification of $\hat{g}_{\mu\nu}$ as the physical metric that
provides the arena on which matter fields interact.  The
matter action, $S_{\mathrm{M}}[\psi^I] =
S_{\mathrm{M}}[\hat{g},\psi^I]$, where $\psi^I$ represents all the
matter fields in spacetime, is one of the standard forms, and
therefore the energy-momentum tensor derived from it is given by
\begin{equation}
\label{eq:matterEM}
\frac{\delta S_{\mathrm{M}}}{\delta
\hat{g}_{\mu\nu}}=-\frac{1}{2}\sqrt{-\hat{g}}\hat{T}^{\mu\nu}.
\end{equation}
It satisfies the conservation laws
\begin{equation}
\label{matterconservation}
\hat{\nabla}_\nu\Bigl[\sqrt{-\hat{g}}\hat{T}^{\mu\nu}\Bigr]=0,
\end{equation}
as a consequence of the matter field equations only.  Note that it
is the covariant derivative $\hat{\nabla}_\mu$ which is compatible
with the metric $\hat{g}_{\mu\nu}$ which appears, and \textit{not}
$\nabla_\mu$ which is defined to be compatible with $g_{\mu\nu}$.
We have included the factor $\sqrt{-\hat{g}}$
in~(\ref{matterconservation}) since this will be a convenient
starting point to derive the consistency of the Bianchi
identities with the field equations.

As an explicit example, if the matter model consisted of a
Maxwell one-form field, then we would have:
\begin{equation}
\label{eq:Maxwell Action} %
S_M=-\int d^4x\sqrt{-\hat{g}}\Bigl[\frac{1}{4}
\hat{g}^{\mu\nu}\hat{g}^{\alpha\beta}
F_{\mu\alpha}F_{\nu\beta}\Bigr],
\end{equation}
where $F_{\mu\nu}=\partial_\mu A_\nu-\partial_\nu A_\mu$.  Note
that we have assumed that it is the density $\sqrt{-\hat{g}}$
that appears in the action, which implies that the
energy-momentum tensor
\begin{equation}
\hat{T}^{\alpha\beta}=\hat{F}^{\alpha\mu}{{}\hat{F}^\beta}_\mu-\frac{1}{4}\hat{F}^2\hat{g}^{\alpha\beta},
\end{equation}
where for example
$\hat{F}^{\alpha\beta}=\hat{g}^{\alpha\mu}\hat{g}^{\beta\nu}F_{\mu\nu}$,
satisfies~(\ref{matterconservation}) if $A_\mu$ satisfies the
field equations $\hat{\nabla}_\beta\hat{F}^{\alpha\beta}=0$.

Since it is only $\hat{g}_{\mu\nu}$ which is `visible' to matter
fields, it is a reasonable assumption (and in line with the
identical assumption implicitly made in general relativity) that
material test bodies will follow geodesics of $\hat{g}_{\mu\nu}$:
\begin{equation}
\frac{d{\hat u}^\alpha}{d\lambda}+{\hat\Gamma}^\alpha_{\mu\nu}{\hat u}^\mu
{\hat u}^\nu=0,
\end{equation}
where $\lambda$ is an affine parameter and the tangent vector ${\hat
u}^\mu$ is normalized as: $\hat{g}_{\mu\nu}{\hat u}^\mu{\hat u}^\nu = 0$,
$\hat{g}_{\mu\nu}{\hat u}^\mu {\hat u}^\nu = c^2$ for null and time-like
geodesics, respectively.

Because all matter fields will couple to $\hat{g}_{\mu\nu}$ in the
same manner, the Weak Equivalence Principle is not violated.
We could easily introduce weak equivalence principle
violating terms into the theory through Yukawa couplings between
the scalar field $\phi$ and the matter fields.  This we leave for
future consideration. Because one can always work in a
locally defined frame with $\hat{g}_{\mu\nu}\approx
\eta_{\mu\nu}$, where $\eta_{\mu\nu}$ is the Minkowski flat
spacetime metric tensor, the field equations for the matter
fields can take on their special relativistic form and the
Einstein Equivalence Principle is not violated, either. However,
if one considers expanding the matter and gravitational fields in
some region of spacetime where $\hat{g}_{\mu\nu}\approx
\eta_{\mu\nu}$, then the perturbation equations for
$\hat{g}_{\mu\nu}$ and $\phi$ will in general \textit{not} take
on their special relativistic form, and therefore the Strong
Equivalence Principle is expected to be violated.

One key feature of our model is that there are no instabilities
induced by this coupling, in the sense of higher-order
derivatives or couplings to unhealthy gauge modes.  To see this we
derive the field equations.

From the variation of the metric~(\ref{bimetric}) we get
\begin{equation}
\delta{\hat g}_{\mu\nu}=\delta g_{\mu\nu}
+2B\partial_{(\mu}\phi\partial_{\nu)}\delta\phi,
\end{equation}
and using the definition~(\ref{eq:matterEM}), we obtain the field
equations
\begin{equation}
\label{fieldeq1}
G^{\mu\nu}=
\frac{\kappa}{2}(T_\phi^{\mu\nu}+s{\hat T}^{\mu\nu}),
\end{equation}
\begin{equation}
\label{fieldeq2} %
\nabla^2\phi+V'[\phi]-\kappa s B{\hat T}^{\mu\nu}
{\hat\nabla}_\mu {\hat\nabla}_\nu\phi =0,
\end{equation}
where $G^{\mu\nu}=R^{\mu\nu}-\frac{1}{2}g^{\mu\nu}R$, $s
=\sqrt{-{\hat g}}/\sqrt{-g}$ and
$\nabla^2\phi=g^{\mu\nu}\nabla_\mu\nabla_\nu\phi$.

As a check that we have indeed produced sensible field equations,
we can show that the Bianchi identities on the curvature of
$g_{\mu\nu}$ are compatible with the field equations. From the
definition~(\ref{bimetric}) we can determine the inverses
\begin{equation}
{\hat g}^{\mu\nu}=g^{\mu\nu}
-\frac{B}{I}\nabla^\mu\phi\nabla^\nu\phi,
\end{equation} and
\begin{equation}
g^{\mu\nu}={\hat g}^{\mu\nu}+\frac{B
}{K}{\hat\nabla}^\mu\phi{\hat\nabla}^\nu\phi,
\end{equation}
where
\begin{equation}
I=1+Bg^{\mu\nu}\partial_\mu\phi\partial_\nu\phi,\quad K=1-B {\hat
g}^{\mu\nu}\partial_\mu\phi\partial_\nu\phi,
\end{equation}
from which it follows that $IK=1$, and we have defined
$\nabla^\mu\phi=g^{\mu\nu}\partial_\nu\phi$ and noted that
\begin{equation}\label{eq:unhat}
{\hat\nabla}^\mu\phi={\hat g}^{\mu\nu}\partial_\nu\phi %
= K\nabla^\mu\phi.
\end{equation}
Using these in the definition of the metric compatible connection
coefficients $\hat{\Gamma}^\alpha_{\mu\nu}$, we find the
equivalent forms
\begin{equation}
\label{gammahat} %
{\hat\Gamma}^\alpha_{\mu\nu}-\Gamma^\alpha_{\mu\nu}
=\frac{B}{I}\nabla^\alpha\phi\nabla_\mu\nabla_\nu\phi %
=\frac{B}{K}\hat{\nabla}^\alpha\phi\hat{\nabla}_\mu\hat{\nabla}_\nu\phi.
\end{equation}

The matter energy-momentum conservation
laws~(\ref{matterconservation}) can be re-written as
\begin{equation}
{\hat\nabla}_\nu \Bigl[\sqrt{-{\hat g}}{\hat T}^{\mu\nu}\Bigr]=
\nabla_\nu\Bigl[\sqrt{-{\hat g}}{\hat
T}^{\mu\nu}\Bigr]+({\hat\Gamma}^\mu_{\alpha\beta}
-\Gamma^\mu_{\alpha\beta})\sqrt{-{\hat g}}{\hat
T}^{\alpha\beta}=0,
\end{equation}
which using~(\ref{gammahat}), become
\begin{equation}\label{eq:DT1} %
{\hat\nabla}_\nu \Bigl[\sqrt{-{\hat g}}{\hat T}^{\mu\nu}\Bigr]= %
\nabla_\nu\Bigl[\sqrt{-{\hat g}}{\hat T}^{\mu\nu}\Bigr] %
+\frac{B}{K}\hat{\nabla}^\mu\phi \sqrt{-{\hat g}}{\hat
T}^{\alpha\beta}\hat{\nabla}_\alpha\hat{\nabla}_\beta\phi=0.
\end{equation}
From the $\nabla_\nu$-derivative of~(\ref{phiT}) we find
\begin{equation}\label{eq:DT2} %
\nabla_\nu[ \sqrt{-g}
T_\phi^{\mu\nu}]=\frac{1}{\kappa}\sqrt{-g}(\nabla^2\phi+V'[\phi])\nabla^\mu\phi %
=\sqrt{-\hat{g}}\frac{B}{K}\hat{\nabla}^\mu\phi\hat{T}^{\alpha\beta}\hat{\nabla}_\alpha\hat{\nabla}_\beta\phi,
\end{equation}
where we have used~(\ref{eq:unhat}) and the scalar field
equations~(\ref{fieldeq2}). Finally, taking the
$\nabla_\nu$-derivative of ($\sqrt{-g}$ times) the field
equations~(\ref{fieldeq1}) we find
\begin{equation}
\nabla_\nu[\sqrt{-g} G^{\mu\nu}] = %
\frac{\kappa}{2}\Bigl(\nabla_\nu [\sqrt{-g} T^{\mu\nu}_\phi] %
+\nabla_\nu [s\sqrt{-g} \hat{T}^{\mu\nu}]\Bigr),
\end{equation}
the right hand side of which vanishes using~(\ref{eq:DT1})
and~(\ref{eq:DT2}).  We have therefore shown
that~(\ref{matterconservation}) is sufficient to guarantee that
the field equations~(\ref{fieldeq1}) are consistent with the
Bianchi identities.

Although we have introduced two metrics into the action, the field
equation for $\phi$ (equation~(\ref{fieldeq2})) is forcing upon
us another null cone.  In order to identify it, we
use~(\ref{gammahat}) to find
\begin{equation}
\hat{\nabla}_\mu\hat{\nabla}_\nu\phi=\frac{1}{I}\nabla_\mu\nabla_\nu\phi,
\end{equation}
which gives the equivalent form of~(\ref{fieldeq2}):
\begin{equation}
\label{equiv1} \biggl(g^{\mu\nu}- s \kappa\frac{B}{I} {\hat
T}^{\mu\nu}\biggr)\nabla_\mu\phi\nabla_\nu\phi+V'[\phi]=0.
\end{equation}
Therefore, there are three light cones in our gravitational theory.
The light cone, $ds^2=g_{\mu\nu}dx^\mu dx^\nu=0$, which describes
the null surfaces of the gravitational propagation, $d{\hat
s}^2=0$ which describes the propagation of ordinary matter, and
$d{\bar s}^2=0$ determined from the metric
\begin{equation}\label{metric3}
{\bar g}^{\mu\nu}=g^{\mu\nu}
- s \kappa \frac{B}{I}{\hat T}^{\mu\nu},
\end{equation}
which describes the propagation of the scalar field $\phi$
obtained from the principal part of~(\ref{equiv1}). There are,
of course, expressions for these in terms of the metric
$\hat{g}_{\mu\nu}$, but they will not be required in their general
form here.

We stress that the field equations do not ``suffer'' from any higher-order
derivative instabilities, and the causality `violation' is
expressed in this class of models as multiple light cones
dynamically determined indirectly by the matter fields. Thus it
can be said that it is a reasonable model with a ``variable speed
of light''.

\section{Cosmological Model}

We shall assume that spacetime and the scalar field $\phi$ are
homogeneous and isotropic.  We will also begin by writing the
metric $g_{\mu\nu}$ in comoving form, which leads us to the
standard Friedmann-Robertson-Walker (FRW) metric:
\begin{equation}
\label{metric1}
ds^2
=c^2dt^2-R^2(t)\biggl[\frac{dr^2}{1-kr^2}+r^2d\theta^2+r^2\sin^2\theta
d\phi^2\biggr],
\end{equation}
where we employ a dimensionless radial coordinate $r$ and $k=0,\pm 1$ for
the flat, closed and hyperbolic spatial topologies, respectively. Let us
write the metric as
\begin{equation}\label{eq:FRWcomoving}
ds^2=c^2dt^2-R^2(t)\gamma_{ij}dx^idx^j.
\end{equation}
For the metric ${\hat g}_{\mu\nu}$ we obtain
\begin{equation}
\label{metric2} d{\hat s}^2
=I(t)c^2dt^2-R^2(t)\gamma_{ij}dx^idx^j,
\end{equation}
where
\begin{equation}
I(t)=1+\frac{B}{c^2}{\dot\phi}^2(t),
\end{equation}
and ${\dot\phi}=d\phi/dt$. We have that $s=\sqrt{I}$ and that $I >
0$ to guarantee that the metric $\hat{g}_{\mu\nu}$ is real and
non-degenerate. From~(\ref{metric3}), we find
\begin{equation}\label{FRWmetric3}
d{\bar s}^2
=\biggl(1-\frac{c^2\kappa
B}{I^{3/2}}\rho_M\biggr)c^2dt^2 - \biggl(1+\frac{\kappa
B}{I^{1/2}}p_M\biggr)R^2(t) \gamma_{ij}dx^idx^j,
\end{equation}
and so we have that $I^{3/2}>c^2\kappa B\rho_M$ for this metric
to be non-degenerate.

Let us assume a perfect fluid model for the energy-momentum tensor
\begin{equation}
{\hat T}^{\mu\nu}=(\rho_M+\frac{p_M}{c^2}){\hat u}^\mu{\hat
u}^\nu -p_M{\hat g}^{\mu\nu},
\end{equation}
where ${\hat u}^\mu=dx^\mu/d{\hat s}$ is normalized to ${\hat g}_{\mu\nu}
{\hat u}^\mu{\hat u}^\nu=c^2$, so that the only non-vanishing component is
${\hat u}^0=c/\sqrt{{\hat g}_{00}}=c\sqrt{{\hat g}^{00}}$. We obtain
\begin{equation}
{\hat T}^{00}=\frac{\rho_M}{I},\quad {\hat T}^{0i}=0,\quad {\hat
T}^{ij}=\frac{p_M}{R^2}\gamma^{ij},
\end{equation}
and from~(\ref{matterconservation}) the matter conservation
equation:
\begin{equation}
\label{conserve}
{\dot\rho_M}+3\biggl(\rho_M+\frac{p_M}{c^2}\biggr)\frac{{\dot R}}{R}=0.
\end{equation}

The field equations~(\ref{fieldeq1}) become
\begin{equation}
\label{equation1}
\biggl(\frac{{\dot R}}{R}\biggr)^2+\frac{c^2k}{R^2}
=\frac{1}{3}c^2\Lambda+\frac{1}{6}\biggl(\frac{1}{2}{\dot\phi}^2
+c^2V[\phi]\biggr)+\frac{\kappa c^4}{6}\frac{\rho_M}{\sqrt{I}},
\end{equation}
\begin{equation}
\label{equation2} \biggl(\frac{{\dot R}}{R}\biggr)^2+2\frac{{\ddot
R}}{R}+\frac{c^2k}{R^2}
=c^2\Lambda-\frac{1}{2}\biggl(\frac{1}{2}{\dot\phi}^2-c^2V[\phi]\biggr)
-\frac{\kappa c^2}{2}{\sqrt{I}}p_M,
\end{equation}
and the scalar field equation~(\ref{fieldeq2}) reduces to
\begin{equation}
\label{ddotphi}
\frac{1}{c^2}\biggl(1-\frac{c^2\kappa B}{I^{3/2}}\rho_M\biggr){\ddot\phi}
+\frac{3{\dot R}}{c^2R}{\dot\phi}\biggl(1+\frac{\kappa B}
{\sqrt{I}}p_M\biggr)+V'[\phi]=0.
\end{equation}

The field equations~(\ref{equation1}-\ref{ddotphi}) are written in
a comoving frame for the gravitational metric $g_{\mu\nu}$.  In
order to make a more obvious connection with standard cosmology,
we shall transform our field equations by using the time
coordinate defined by
\begin{equation}
\tau=\int\sqrt{I}dt,
\end{equation}
which puts the matter metric ${\hat g}_{\mu\nu}$ into comoving
coordinate form.  In this new frame, the field equations become
\begin{equation}
\label{coeq1}
\biggl(\frac{{\dot R}}{R}\biggr)^2+\frac{c^2kK}{R^2}
=\frac{1}{3}c^2\Lambda K+\frac{1}{6}
\biggl(\frac{1}{2}{\dot\phi}^2+c^2KV[\phi]\biggr)
+\frac{1}{6}\kappa c^4K^{3/2}\rho_M,
\end{equation}
\begin{equation}
\label{coeq2}
\biggl(\frac{{\dot R}}{R}\biggr)^2+2\frac{{\ddot
R}}{R}+\frac{c^2kK}{R^2} =\frac{{\dot K}{\dot
R}}{KR}+c^2\Lambda K-\frac{1}{2}\biggl(\frac{1}{2}{\dot\phi}^2
-c^2KV[\phi]\biggr)-\frac{1}{2}\kappa c^2\sqrt{K}p_M,
\end{equation}
and
\begin{equation}\label{scalarco}
\frac{1}{c^2}\biggl(1-\kappa c^2B K^{3/2}\rho_M\biggr){\ddot\phi}
+\frac{3K{\dot R}}{c^2R}{\dot\phi}\biggl(1+ \kappa
B\sqrt{K}p_M\biggr)+K^2V'[\phi]=0.
\end{equation}

Making the definitions
\begin{equation}
\label{scalar} \rho_\phi=\frac{1}{\kappa
c^2}\biggl(\frac{1}{2c^2}{\dot\phi}^2+KV[\phi]\biggr),\quad
p_\phi=\frac{1}{\kappa}\biggl(\frac{1}{2c^2}{\dot\phi}^2-KV[\phi]\biggr),
\end{equation}
we see that $K=1-\kappa c^2 B(\rho_\phi+p_\phi/c^2)$, and
that $\rho_B:=1/(\kappa c^2 B)$ gives the energy scale at which
$K$ deviates significantly from $1$, and therefore significant
deviations from a standard general relativity plus matter model are
to be expected.

Defining: $\rho_\Lambda=2\Lambda/(c^2\kappa)$,
$p_\Lambda=-2\Lambda/\kappa$, $\Omega_k=kc^2/(R^2H^2)$,
$\Omega_\Lambda=c^4\kappa\rho_\Lambda/(6H^2)$,
$\Omega_\phi=c^4\kappa\rho_\phi/(6H^2)$ and
$\Omega_M=c^4\kappa\rho_M/(6H^2)$, we can write the Friedmann
equation~(\ref{coeq1}) in the ``sum-rule'' form:
\begin{equation}\label{Friedmann}
1+K\Omega_k=K\Omega_\Lambda+\Omega_\phi+K^{3/2}\Omega_M.
\end{equation}
Although this form is similar to that
which appears in standard cosmological models, the factor $K$ and
$\Omega_\phi$ depend on the scalar field
and therefore this is \textit{not} just a simple sum of individual
energy contributions. From~(\ref{coeq2}) we can derive an
expression for the deceleration parameter:
\begin{equation}
q=-\frac{\ddot{R}}{H^2R} %
=-\frac{\dot{K}}{2HK} +\frac{1}{2}(1+K\Omega_k)
+\frac{c^2\kappa}{4H^2}(p_\phi+K^{1/2}p_M+Kp_\Lambda).
\end{equation}
The first term can be evaluated by using the scalar field
equation~(\ref{scalarco}) to give
\begin{equation}
\frac{\dot{K}}{2HK}=\frac{3(1-K)(1+\kappa B
K^{1/2}p_M)+H^{-1}B\dot{\phi}KV^\prime[\phi]}{1-\kappa c^2 B
K^{3/2}\rho_M}.
\end{equation}
Note that the denominator is positive-definite wherever the
metric~(\ref{FRWmetric3}) has the correct signature, the first
term in the numerator is positive-definite whenever the
metric~(\ref{metric1}) has the correct signature, and the second
term in the numerator is positive provided that the universe is
expanding $H>0$ and $\partial_t V[\phi]>0$.

\section{Parameterization of the Cosmological Model}

We shall now consider the present universe modeled by a spatially
flat FRW spacetime containing non-relativistic matter and the
scalar field $\phi$. We shall assume the standard equation of
state with negligible matter pressure $p_M\approx 0$, so that from
the conservation equation~(\ref{conserve}) we get
\begin{equation}
\frac{\rho_M}{\rho_{0\,M}}=\biggl(\frac{R_0}{R}\biggr)^3,
\end{equation}
where $\rho_{0\,M}=\rho_M(t_0)$, $R_0=R(t_0)$
denote the present values of the density and cosmic scale factor
with $t_0$ denoting the present value of time $t$. Taking
$k=\Lambda=0$, Eq.~(\ref{Friedmann}) becomes
\begin{equation}\label{Friedmannred}
1=\Omega_{0\,\phi}+K_0^{3/2}\Omega_{0\,M}.
\end{equation}
We can obtain a fit to the present data by essentially following
the method of Perlmutter et al.~\cite{Perlmutter} and choosing for
a spatially flat universe the best fit value: $\Omega_{0\,M}=0.28$
and the parameterization:
\begin{equation}
\Omega_{0\,\phi}=1-0.28K_0^{3/2}.
\end{equation}

The present value of the deceleration parameter can be
written
\begin{equation}
q_0=-\frac{\dot{K}_0}{2H_0K_0} +\frac{1}{2}
+\frac{c^2\kappa}{4H_0^2}p_{0\,\phi}.
\end{equation}
We see that in our gravity theory, we can achieve a negative
deceleration parameter $q_0$, if the first term on the right-hand
side dominates. We stress that the cosmic acceleration with
${\ddot R}(t) > 0$ is {\it caused by the dynamics of the scalar
field} $\phi$, which acts as a gravitational field component, and
the matter pressure $p_M$ can be negligible and the cosmological
constant $\Lambda$ can be zero. Galaxy formation at earlier times
can be obtained for small and positive $q$, if the solution for
$\phi$ and the pressure $p_\phi$ are appropriately chosen. As has
been demonstrated elsewhere, the standard horizon and flatness
problems can also be resolved by superluminary
models~\cite{Vector,Superlum1,Albrecht,Superlum2}. In the present
theory, an early universe inflationary expansion phase can be
obtained by a suitable choice of initial conditions for $\phi$
and $V(\phi)$.  A solution that accomplishes this has been found
and will be presented in a forthcoming article.

\section{Experimental Signature for Gravitational Waves}

Let us define
\begin{equation}
{\bar c}=c\sqrt{K}=c(1-\frac{B}{c^2}{\dot\phi}^2)^{1/2},\quad
{\bar G}=GK^{3/2}=G(1-\frac{B}{c^2}{\dot\phi}^2)^{3/2},
\end{equation}
and
\begin{equation}
\bar{\Lambda}=\Lambda+\frac{1}{2} %
\Bigl(\frac{1}{2\bar{c}^2}\dot{\phi}^2+V[\phi]\Bigr),
\end{equation}
so that we can rewrite~(\ref{coeq1}) as
\begin{equation}\label{eq:VPC}
H^2+\frac{{\bar c}^2k}{R^2} =\frac{1}{3}{\bar c}^2\bar{\Lambda}
+\frac{8\pi {\bar G}}{3}\rho_M.
\end{equation}
This has the form of the Friedmann equation in Einstein gravity
with an ``effective'' velocity of light ${\bar c}$, gravitational
constant ${\bar G}$ and cosmological constant $\bar{\Lambda}$. We
have therefore mapped our model to a particular case of the
models considered in~\cite{Albrecht}.  It is interesting to note
that in this particular model
$\bar{G}/\bar{c}^3=G/c^3=\textrm{constant}$.

These effective time varying constants are merely definitions that
allow us to write the Friedman equation in the standard form.
Although $\bar{G}$ may be interpreted as an effective
gravitational coupling to matter, we must be careful how we
interpret $\bar{c}$ as a time varying speed of light. In fact, in
this frame matter field perturbations will propagate at a speed
determined by the line element $d\hat{s}^2=c^2
dt^2-R^2(t)\gamma_{ij}dx^idx^i$, whereas metric perturbations 
will propagate at a speed determined by $ds^2=\bar{c}^2
dt^2-R^2(t)\gamma_{ij}dx^idx^i$; in both cases $R(t)$ is the
solution of~(\ref{eq:VPC}). This means that matter fields will
behave as in conventional models, but the gravitational
field cannot be understood in terms of a minimal coupling to
these fields.  Note that this rules out effects like those
considered in~\cite{Barrow+Magueijo:1999}.

We can now predict that there will be a time lag $\Delta t$ between the
speed of gravitational waves travelling along the null surface of the
gravitational light cone $ds^2=0$ and photons and other matter particles
travelling through and on the null surface of the matter light
cone $d{\hat s}^2=0$. This could be the basis of an important observational
signature for our gravity theory, for with $K\approx 0$ in the early
universe the time lag $\Delta t$ could be a measurable quantity in
gravitational wave experiments. Morover, the effective gravitational
constant ${\bar G}$ would by the same reasoning be smaller in the early
universe than its presently measured value $G$.

\section{Conclusions}

We have developed a gravity theory based on a bimetric structure.
When $B=0$, we regain standard GR. There are three light cones
associated with the metrics $g_{\mu\nu}$, ${\hat g}_{\mu\nu}$ and
${\bar g}_{\mu\nu}$. All matter fields except the scalar field
$\phi$ propagate in the geometry described by ${\hat g}_{\mu\nu}$
and material test particles and photons will propagate along
geodesics determined by this geometry and obey the equivalence
principle. In future work the properties and predictions of
static spherically symmetric solutions for gravitational
phenomena will be investigated. The theory could have significance
for the problem of singularities and the physical properties of
black holes. There is an interesting experimental signature in
our gravity theory, due to the slowing down of gravitational
waves emitted in the early universe which could be a detectable
physical phenomenon in gravitational wave experiments.

The cosmological model is in agreement with the magnitude-redshift relation
of Type Ia supernovae for $\Omega_{0\,M}=0.28$. In a recent paper,
Caldwell~\cite{Caldwell} considered the possibility that future
observational constraints and more accurate supernovae data may require
that quintessence models may need significantly more negative values for
the pressure in the equation of state. In our model this would correspond
to an increase in the size of $\Omega_{0\,\phi}$ keeping the density
positive and $p_M\approx 0$. 

\vskip 0.2 true in
{\bf Acknowledgments}
\vskip 0.2 true in
J. W. Moffat was supported by the Natural Sciences and Engineering Research
Council of Canada.
\vskip 0.5 true in

\end{document}